\documentclass[showpacs,preprintnumbers,amsmath,amssymb,pre,floatfix,showkeys]{revtex4}

\usepackage[english]{babel}
\usepackage[latin2]{inputenc}
\usepackage[T1]{fontenc}

\usepackage[usenames]{color}

\usepackage{ae,aecompl}
\usepackage{enumerate}
\usepackage{epsfig}
\usepackage{graphics}
\usepackage{graphicx}
\usepackage{amsmath}
\usepackage{amssymb}
\usepackage{pifont}
\usepackage[sort&compress]{natbib}
\usepackage{wasysym}
\usepackage{color}

\def\ev #1{\left\langle #1 \right\rangle}

\begin{document}

\preprint{}
\title{Volatility: a hidden Markov process in financial time series}
\author{Zolt\'an Eisler}
\email{eisler@maxwell.phy.bme.hu}
\affiliation{Department of Theoretical Physics, Budapest University of Technology and Economics \\ Budafoki \'ut 8., H-1111, Budapest, Hungary}
\author{Josep Perell\'o}
\email{josep.perello@ub.edu}
\author{Jaume Masoliver}
\email{jaume.masoliver@ub.edu}
\affiliation{Departament de F\'{\i}sica Fonamental, Universitat de
Barcelona\\ Diagonal 647, E-08028 Barcelona, Spain}


\date{\today}

\begin{abstract}
The volatility characterizes the amplitude of price return fluctuations. It is a central magnitude in finance closely related to the risk of holding a certain asset. Despite its popularity on trading floors, the volatility is unobservable and only the price is known. Diffusion theory has many common points with the research on volatility, the key of the analogy being that volatility is the time-dependent diffusion coefficient of the random walk for the price return. We present a formal procedure to extract volatility from price data, by assuming that it is described by a hidden Markov process which together with the price form a two-dimensional diffusion process. We derive a maximum likelihood estimate of the volatility path valid for a wide class of two-dimensional diffusion processes. The choice of the exponential Ornstein-Uhlenbeck (expOU) stochastic volatility model performs remarkably well in inferring the hidden state of volatility. The formalism is applied to the Dow Jones index. The main results are: (i) the distribution of estimated volatility is lognormal, which is consistent with the expOU model and (ii) the estimated volatility is related to trading volume by a power law of the form $\sigma \propto V^{0.55}$; and (iii) future returns are proportional to the current volatility which suggests some degree of predictability for the size of future returns.
\end{abstract}

\maketitle

\section{Introduction}
\label{sec:intro}

The volatility measures the amplitude of return fluctuations and it is one of the central quantities in finance \cite{bouchaud.book}. Investors sometimes place even greater emphasis on the level of volatility than on the market trend itself. The reason for this is mainly that the risk of holding an asset is classically associated with its volatility \cite{markowitz.portfolio}. The theoretical framework used to quantify aspects of price fluctuations has many common points with areas of physics dealing with noisy signals. The research of random diffusion aims to describe the dynamics of particles in random media and its methods have been applied to a large variety of phenomena in statistical physics and condensed matter \cite{ben}. Time series describing solar flares, earthquakes, the human heartbeat or climate records show strong correlations, multi-scaling, non-Gaussian statistics and self-organized behavior  \cite{paczuski.solar,bak.unified,bunde.heartbeat,bunde.climate}. These are properties also observed in financial time series where volatility is considered to play a key role \cite{bouchaud.book,cont}.

The picture that prices follow a simple diffusion process was first proposed by Bachelier in 1900 \cite{bachelier}. Later in 1959, the physicist Osborne introduced the geometric Brownian motion and suggested that volatility can be viewed as the diffusion coefficient of this random walk \cite{osborne}. The simplest possible assumption -- that it is a time-independent constant -- lies at the heart of classical models such as the Black-Scholes option pricing formula \cite{black.scholes}. More recently it has become widely accepted that such an assumption is inadequate to explain the richness of the markets' behavior \cite{bouchaud.book}. Instead, volatility itself should be treated as a random quantity with its own particular dynamics.

Among its most relevant properties \cite{bouchaud.book,cont,lo,ding,stanley}, volatility is the responsible for the observed clustering in price changes. That is: large fluctuations are commonly followed by other large fluctuations and similarly for small changes \cite{bouchaud.book,cont}. Another related feature is that, in clear contrast with price changes which show negligible autocorrelations, volatility autocorrelation is still significant for time lags longer than a year \cite{cont,lo,ding}. Most of these studies introduce a subordinated process which is associated with the volatility in one way or another \cite{bns,saichev,bacry,fisher,lux,calvet.estimation,masoliver.qf,ronnie.book,pm,pm2,perello2,yakovenko}. 

The main obstacle of the appropriate analysis of volatility is that it is directly unobservable. As we have mentioned, volatility provides important information to traders but it is very unclear how reliable the estimates of such a hidden process can be. Investors use several proxies to infer the level of current asset volatility. The most common ways are: (i) to make it equivalent to the absolute value of return changes, (ii) to assume a proportional law between volatility and market volume \cite{yakovenko,andersen,brock,lindberg}, and (iii) to use the information contained in  option prices and obtain the so-called ``implied volatility'' which, in fact, corresponds to \emph{the market's belief of volatility} \cite{bouchaud.book}. 

This already raises the questions: What process is a proper model of volatility and how to adjust the possible parameters to describe various stocks and markets? Among the possible candidates, multifractals \cite{calvet.estimation,saichev,bacry,fisher,lux} and stochastic volatility models \cite{masoliver.qf,ronnie.book,pm,pm2,perello2,yakovenko,bns} are the most promising. The fact that volatility is directly unobservable makes even more difficult to get a conclusive answer about the best model. This doubled challenge has deserved the attention from the most diverse disciplines that look at financial markets. All of them have converged on the description of what is technically known as realized volatility and on methodologies for reconstructing volatility path. Several procedures for reconstructing volatility have already been presented being more or less dependent on the volatility model chosen. Mathematics, econometrics, and finance research have a large number of publications during the last decade on this issue (see e.g. \cite{bns,andersen2,genon,elerian,bns2,barucci,genon2,genon3,griffin,jong,morimune,omori,reno}).

Our research presents an alternative procedure to estimate volatility from the price dynamics only. Likewise most of the papers in the literature, the price dynamics is represented by a two-dimensional diffusion process: one dimension for price and second dimension for a volatility described by a hidden Markov process. Our case estimates the volatility subordinated time series through maximum likelihood but, in contrast with other studies, having fixed in advance the parameters of the model.

We have decided to focus on a particular stochastic volatility model that is able to circumvent both mathematical and computational difficulties: the exponential Ornstein-Uhlenbeck volatility model \cite{masoliver.qf,ronnie.book} although the same procedure can be used for a larger class of models. As its name indicates, the model assumes that the logarithm of the volatility follows an Ornstein-Uhlenbeck process, that is: a mean reverting process with linear drift. The resulting model is capable of reproducing the statistical properties of the financial markets fairly well~\cite{masoliver.qf}. 

The paper is organized as follows. In Sec. \ref{sec:sv} we outline the general stochastic volatility framework and more specifically the exponential Ornstein-Uhlenbeck model. In Sec. \ref{sec:est} we present a maximum likelihood estimator for a wide class of stochastic volatility models. For the case of the exponential Ornstein-Uhlenbeck (expOU) model we present Monte Carlo simulations to show that it performs remarkably well in inferring the hidden state of the volatility process. In Sec. \ref{sec:app} the procedure is applied to the Dow Jones Industrial Average. Conclusions are drawn in Sec. \ref{sec:conc} and some more technical details are left to the appendices.

\section{Stochastic volatility models}
\label{sec:sv}

The geometric Brownian motion (GBM)~\cite{osborne} is the most widely used model in finance. In this setting the asset price $S(t)$ is described through the following Langevin equation (in It\^o sense):
\begin{equation}
\frac{dS(t)}{S(t)}=\mu dt+\sigma dW_1(t),
\label{1}
\end{equation}
where $\sigma$ is the volatility, assumed to be constant, $\mu$ is some deterministic drift indicating an eventual trend in the market, and $W_1(t)$ is the Wiener process. We define the zero-mean return $X(t)$ as
\begin{equation}
X(t)=\ln\left[S(t+t_0)/S(t_0)\right]-\left\langle\ln\left[S(t+t_0)/S(t_0)\right]\right\rangle,
\label{return}
\end{equation}
where the symbol $\langle\cdots\rangle$ designates the average value and $t_0$ is the initial time which is usually set to be zero. In terms of $X(t)$ the GBM is simply written as
\begin{equation}
dX(t)=\sigma dW_1(t).
\label{1r}
\end{equation}

However, especially after the 1987 market crash, compelling empirical evidence has become available that the assumption of a constant volatility is doubtful~\cite{cont}. Neither is volatility a deterministic function of time as one might expect on account of the non-stationarity of financial data, but a random quantity \cite{ronnie.book}. 

In the most general setting one therefore assumes that the volatility $\sigma$ is a given function of a random process $Y(t)$: 
\begin{equation}
\sigma(t)=f[Y(t)].
\label{3}
\end{equation}
Most stochastic volatility (SV) models assume $Y(t)$ is also a diffusion process that may or may not be correlated with price. The main difference between various models is only the parametrization of this scheme. In a general notation the zero-mean return $X(t)$ defined above is described by the following set of stochastic differential equations
\begin{eqnarray}
dX(t) &=& f[Y(t)]dW_1(t)
\label{eq:sv1} \\
dY(t) &=& -g[Y(t)]dt+h[Y(t)]dW_2(t),
\label{eq:sv2}
\end{eqnarray}
where $dX=dS/S-\langle dS/S\rangle$ and $f$, $g$ and $h$ are given functions of $Y(t)$. As shown in Eq. (\ref{3}),  $f[Y(t)]$ corresponds to the volatility, i.e., the amplitude of return fluctuations. However, since $f(x)$ is usually chosen to be a monotonically increasing function, it is not misleading to think of $Y$ as a measure of volatility. Thus, as far as there is no confusion, we will refer to the process $Y(t)$ as ``volatility'' as well. On the other hand, the function $g[Y(t)]$ describes a reverting force that drives the volatility toward the so-called ``normal level''. This force brings the volatility process to a stationary regime for long time horizons and the normal level is related to the average volatility in that limit. Finally, the subordinated process $Y(t)$ may have a non-constant diffusion coefficient defined in terms of the function $h[Y(t)]$ which is called the  volatility-of-volatility (``vol of vol"). The functions $g$ and $h$ fully describe the volatility process. The resulting dynamics is comparable with the one described by a Gaussian particle trapped in a potential well $V(y)$ whose associated force is $-g(y)$, where $g(y)=V'(y)$. In finance one typically proposes convex potentials with only one minimum whose value is related to the normal level of the volatility.

In what follows we will mostly work with one particular SV model, the exponential Ornstein-Uhlenbeck (expOU) model,  which follows from the substitutions:
$$
f(x)=me^x, \quad g(x)=\alpha x, \quad h(x)=k, 
$$
that is, 
\begin{eqnarray}
dX(t)&=&me^{Y(t)}dW_1(t),
\label{eq:expou1} \\
dY(t)&=&-\alpha Y(t)dt+kdW_2(t).
\label{eq:expou2}
\end{eqnarray}
Note that in this model the process $Y(t)$ is precisely the logarithm of the volatility, or ``log-volatility'' for short. The main statistical properties of the model are thoroughly discussed in Ref.~\cite{masoliver.qf}. We simply recall that the stationary distribution of the process $Y(t)$ is a Gaussian ({\it i.e.,} a lognormal distribution for $\sigma$) with zero mean and variance $\beta$:
\begin{equation}
p(y)=\frac{1}{\sqrt{2\pi \beta}}\exp\left(-y^2/2\beta\right),
\label{eq:p}
\end{equation}
where 
\begin{equation}
\beta\equiv k^2/2\alpha.
\label{variance}
\end{equation}

\section{Volatility estimation}
\label{sec:est}
\subsection{The Wiener measure and volatility estimation}

Let $\mathbf X$ and $\mathbf Y$ denote a simultaneous realization of the variables $X(\tau)$ and $Y(\tau)$ in the time interval $t-s\leq\tau\leq t$. Omitting all $\mathbf Y$-independent terms, we show in Appendix \ref{app:proof} that the probability density (likelihood) of such a realization is approximately given by
\begin{widetext}

\begin{equation}
\ln \mathbb P(\mathbf X, \mathbf Y)\simeq-\frac{1}{2}\int_{t-s}^t\left[\frac{\dot X(\tau)}{f[Y(\tau)]}\right]^2 d\tau
-\frac{1}{2}\int_{t-s}^t\left[\frac{\dot Y+g[Y(\tau)]}{h[Y(\tau)]}\right]^2 d\tau+\dots
\label{eq:logp}
\end{equation}
\end{widetext}

Before proceeding further, we will discuss the meaning of this expression. We first note that Eq. (\ref{eq:logp}) has to be understood in the sense of generalized functions \cite{gelfand} since the Wiener process is only differentiable in this sense and, hence, $\dot X(t)$ and $\dot Y(t)$ do not exist as ordinary functions and Eq. (\ref{eq:logp}) is just a symbolic expression. Nevertheless, the formula is still valid when the integral and the derivatives therein are discretized with arbitrary small time steps, a requirement that is indeed necessary for numerical computations.

Let us now see some qualitative properties of Eq. (\ref{eq:logp}). The first summand measures the fluctuations of the zero-mean return with respect to the volatility, $[\dot X(\tau)/f(Y(\tau))]^2$, and their contribution to the likelihood (probability density) of a given return realization. Note that the higher this contribution is, the lower those ``relative'' fluctuations are. In the same fashion, the second summand in Eq. (\ref{eq:logp}) measures the fluctuations of the volatility process $Y(t)$ with respect to the vol of vol $h(Y)$, although in this case these fluctuations are gauged with the mean reverting force $-g(Y)$. As before, the lower these fluctuations, the higher their contribution to the log-likelihood (\ref{eq:logp}). 

While Eqs. (\ref{eq:sv1})-(\ref{eq:sv2}) represent a joint model for return and volatility, the stock market data only include recordings of the return process $X$. The $Y$ process and, hence, the volatility $f(Y)$, must be inferred indirectly in a Bayesian fashion through Eq. \eqref{eq:logp}. Indeed, the conditional probability density that the realization of the hidden $Y$-process is $\mathbf Y$, given that the observed return is $\mathbf X$, reads
\begin{eqnarray}
\ln \mathbb P(\mathbf Y\vert \mathbf X)=\ln \mathbb P(\mathbf X, \mathbf Y)-\ln \mathbb P(\mathbf X).
\label{eq:logpcond}
\end{eqnarray}
In consequence, {\it we can find the maximum likelihood sample path of the (hidden) volatility process by maximizing Eq. (\ref{eq:logpcond}) with respect to $\mathbf Y$} (recall that $\mathbf Y$ is a realization of $Y(\tau)$ in the interval $t-s\leq\tau\leq t$). Since the second summand of Eq. (\ref{eq:logpcond}) is independent of $\mathbf Y$, we can neglect $\mathbb P(\mathbf X)$ in this maximization process. Therefore, the maximization of 
$\ln \mathbb P(\mathbf Y\vert \mathbf X)$ yields the same result as that of $\ln \mathbb P(\mathbf X, \mathbf Y)$.

Note that, besides the specification of the stochastic volatility model (that is, the explicit forms of $f,g$, and $h$), the only free parameter is $s$: the duration of past return data to take into account. After substituting the observed return history as $\mathbf X$, we will obtain by maximum likelihood the quantity:
\begin{equation}
\hat {\mathbf Y} = \mathrm{argmax}_{\mathbf Y} \ln \mathbb P(\mathbf Y\vert \mathbf X)= 
\mathrm{argmax}_{\mathbf Y} \ln \mathbb P(\mathbf X,\mathbf Y).
\end{equation}

We should mention that similar maximization problems have been studied in the context of hidden Markov models, where this procedure is called ``decoding''. When the state space (the number of possible $X$ and $Y$ values) is finite, the optimization can be done exactly by the Viterbi algorithm \cite{viterbi}, while there has been limited success in the continuous case \cite{csva}. A similar technique has been applied to the forecasting of volatility assuming a binomial cascade model which, unlike stochastic volatility models, has a finite state space (see Ref. \cite{calvet.estimation}). More focused on the stochastic volatility, we should also mention the efforts based on Kalman or Particle filtering, Bayesian inference, conditional likelihood or Fourier methods among other similar techniques (see e.g.  Refs.~\cite{bns,andersen2,genon,elerian,bns2,barucci,genon2,genon3,griffin,jong,morimune,omori,reno}). Most of the cited works assume an specific model, although they never focuss on the expOU model. In addition, they are mainly worried about intraday (high-frequency) data while we are here focussed on reproducing daily (low-frequency) data. It deserves an special attention the work on a superposition of Lévy Ornstein-Uhlenbeck processes for $\sigma^2$ crafted to describe high-frequency (intraday) data and provide a power law slow decay for the volatility autocorrelation~\cite{bns,bns2}. All the techniques mentioned above provide an efficient way to reproduce the volatility path but, in contrast to our case, the method also serves to estimate the parameters of the model. We have decided to provide the parameters beforehand thus using an independent way of estimating them (see Sec.~\ref{sec:param}). It should be left for a future research an accurate comparison between our method and others. We may look for a way to implement parameter estimation in our method.

As we have already stated, our main objective is to design a method able to filter the Wiener noise $dW_1(t)$ out of Eq. (\ref{eq:sv1}) and thus to obtain a reliable estimate $\hat{Y}(t)$ of the hidden volatility process $Y(t)$. The method, an extension of the deconvolution procedure previously presented in \cite{masoliver.qf}, has basically the following five steps:
\begin{enumerate}[(i)]
\item We simulate a random sample path of the Wiener process $\widehat{d W_1}(\tau)$ for $t-s\leq\tau\leq t$. 
\item Then a surrogate realization of $\mathbf Y$ is generated as
\begin{equation}
\hat Y_s(\tau) = f^{-1}\left(\left\vert\frac{dX(\tau)}{\widehat{d W_1}(\tau)}\right\vert\right),
\label{hatY}
\end{equation}
where $t-s\leq\tau\leq t$. Note that this equation requires that $f(x)$ is invertible which implies that $f(x)$ be chosen to be a monotonic function.
\item Substitute $\hat{\mathbf Y}_s$ and $\mathbf X$ into Eq. \eqref{eq:logp} to calculate the log-likelihood of this realization.
\item Iterate (i)-(iii) for $I$ steps, keep the highest likelihood random realization (the conditional median) and assume this to be the proper estimate $\hat Y(t)$.
\item The estimate of the hidden process at time $t$ is then $\hat Y(t)$. The estimate of the volatility is given by 
\begin{equation}
\hat \sigma(t) = f[\hat Y(t)].
\end{equation}
\end{enumerate}

\subsection{Interpretation of the method}

Let us further elaborate the meaning of such an estimate. In finance, the volatility is often identified with the absolute value of returns variations. Indeed, as a first approximation, we can replace in Eq. (\ref{eq:sv1}) the noise term by its expected value and write
\begin{equation}
\sigma(t) \approx \frac{|dX(t)|}{\ev{|dW_1(t)|}},
\label{eq:dxabs}
\end{equation}
which shows that the volatility is approximately proportional to the absolute returns. Eq. (\ref{eq:sv1}) can be thus thought of as a first approximation toward estimating volatility. Our method, based on the maximization of Eq. (\ref{eq:logpcond}), takes this estimation two steps further. In effect, the first step was taken in Ref. \cite{masoliver.qf} where we replaced the average $\ev{|dW_1(t)|}$ by a simulated sample path. We are now taking a second and more refined step in which we are not only replacing the Wiener noise by a random simulation but, in addition, we perform the maximum likelihood method described by items (i)-(v). 

Thus we are basically separating the observed returns $dX(t)$ into two sources: $\sigma(t)$ and $dW_1(t)$. To do this, we have first considered a specific form of stochastic volatility. Secondly, we have taken the driving Wiener noises $dW_1(t)$ and $dW_2(t)$ appearing in Eqs. (\ref{eq:sv1}) and (\ref{eq:sv2}) to be uncorrelated. Finally, we have assumed that $\sigma(t)$ is approximately constant over the time step during which we numerically evaluate the derivatives $\dot{Y}(t)$ and $\dot{X}(t)$ appearing in Eq. (\ref{eq:logp}). We incidentally note that if $h(x)=0$ -- the vol of vol is equal to zero -- then the stationary solution of Eq. \eqref{eq:sv2} is $Y(t)\equiv 0$. Thus the model reduces to the Wiener process in which the volatility is constant and absolute returns are uncorrelated \cite{bachelier}.

\subsection{The performance of the estimator}
\label{sec-perf}

\begin{figure}[ptb]
\centerline{\includegraphics[height=180pt]{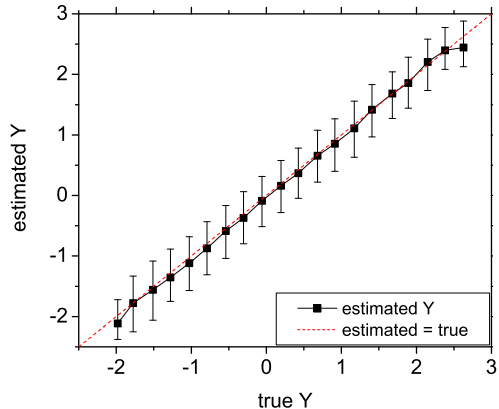}}
\caption{Estimated log-volatility $\hat Y$ as a function of the actual log-volatility $Y$ taken from $2\times 10^5$ simulations of the expOU model. Reconstruction used the last $s=10$ values of returns, and $I=10^5$ iterations. The error bars represent the $25\%$ and $75\%$ quantiles of the distribution estimated volatility.}
\label{fig:reconstruction}
\end{figure}

In order to test the performance of the estimator, we simulate the expOU process by using Eqs. (\ref{eq:expou1})-(\ref{eq:expou2}) with the realistic parameters obtained in Section \ref{sec:app}. The relationship between the simulated value of the log-volatility $Y(t)$ and its estimate 
$\hat Y(t)$ is given in Fig. \ref{fig:reconstruction}. The two quantities agree within error bars, so we may state that
$$
Y(t)\approx\hat Y(t) \qquad \mbox{(in mean square sense)}. 
$$

In what follows we will always use $s=10$ days of past data and $I=10^5$ iterations for maximization. The time step for discretization will be $\Delta t = 1$ day. For random optimization, the necessary number of iterations $I$ grows very fastly, perhaps exponentially, with $s/\Delta t$. The values of $s$ and $\Delta t$ cited here were chosen to keep the task computationally feasible. As for the value of $I$, in this case an increase to $I=10^6$ does not improve the estimates noticeably. The estimates generated with this parameter set have negligible bias and they can efficiently distinguish between low and high volatility periods (see Appendix~\ref{app-proc} for a discussion on the robustness and possible bias of the procedure).

\begin{figure}[ptb]
\centerline{\includegraphics[height=180pt]{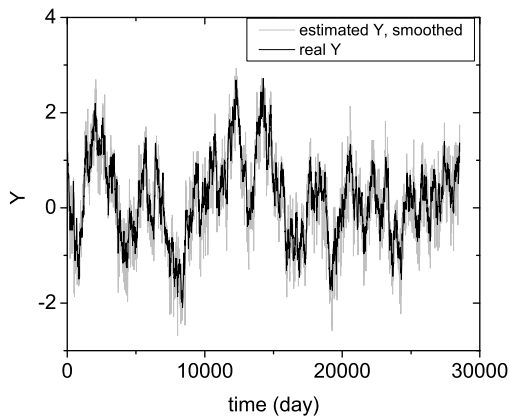}}
\caption{Estimated and actual volatility for a typical sample path of the expOU model. The estimated values were smoothed by $5$-neighbor averaging to reduce noise.}
\label{fig:demonstration}
\end{figure}

An additional verification of the good performance of our estimate is shown in Fig. \ref{fig:demonstration}, where we give the actual sample path of $Y$ for a single realization of the expOU model together with the estimated $\hat Y$. As we can see the estimate follows the true log-volatility $Y(t)$ very closely.

\subsection{Volatility forecasting}

\begin{figure}[ptb]
\centerline{\includegraphics[height=200pt]{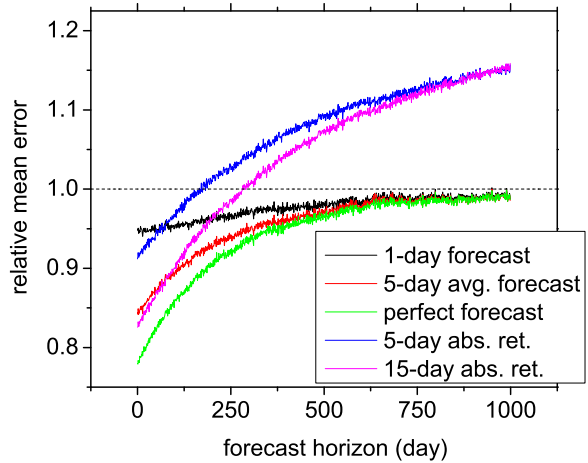}}
\caption{Median forecasting error $E(h)$ of absolute returns calculated according to Eq. \eqref{eq:dxth} by five methods. The errors were normalized by the lowest level of error achievable with the assumption of a time-independent constant volatility (horizontal dashed line). The result was averaged over $150000$ independent realizations of the expOU process, parameters were chosen as outlined in Sec. \ref{sec:param}.}
\label{fig:Gforecastfinalmean}
\end{figure}

Another possible approach to the utility of the expOU model is to evaluate its forecasting performance. One can for example use the mean error of absolute returns $\vert dX(t+h) \vert$ for a given Monte Carlo simulated path:
\begin{eqnarray}
E(h) = \ev{\big\vert\vert \widehat{d X}(t+h) \vert - \vert dX(t+h) \vert \big\vert},
\end{eqnarray}
where $\vert \widehat{d X}(t+h) \vert$ is our estimate for the absolute return $\vert dX(t+h)\vert$ given an estimate $\hat Y(t)$ while $h$ is the time horizon (number of days) to forecast ahead. We recall that for our estimates we take the conditional median of our MonteCarlo simulations. Thus for the expOU model we have
\begin{eqnarray}
\vert \widehat{d X}(t+h) \vert = M[\vert dW_1(t+h)\vert] m\exp[\hat Y(t)\exp(-\alpha h)],
\label{eq:dxth}
\end{eqnarray}
where $M[\vert dW_1(t+h)\vert]$ denotes the median of the absolute value of Wiener increment. Figure \ref{fig:Gforecastfinalmean} compares five methods to forecast $\vert dX(t+h) \vert$ using only information available before time $t$. For clarity we also give the titles used in Fig. 1 in brackets ("..."):
\begin{enumerate}
	\item The median of the last $5$ absolute returns. This method clearly has predictive power for $h < 150$ days. ("5-day abs. ret.")
	\item The median of the last $15$ absolute returns. The longer averaging period brings a substantial improvement for short-term forecasts, but not for long-term.  ("15-day abs. ret.")
	\item Eq. \eqref{eq:dxth} with $\hat Y(t) = Y(t)$, which is the true value in the underlying simulation of the expOU model. This gives a substantial decrease in forecasting error, which persists up to $h \sim 500$ days. This curve is also a theoretical lower bound for error achievable by the expOU model in any time series: Here the underlying data are perfectly described by expOU, and the parameters and $Y(t)$ are known "perfectly". Neither of these usually happen in real data, and so there one expects worse performance. ("perfect")
	\item Eq. \eqref{eq:dxth} with $\hat Y(t)$ estimated by the reconstruction procedure of Sec. \ref{sec:est}. This estimator does not perform well, due to noise in the optimization procedure. ("1-day forecast")
	\item Eq. \eqref{eq:dxth} with $\hat Y(t)$ estimated by the reconstruction procedure of Sec. \ref{sec:est}, then averaged for the last $5$ days. Such averaging greatly decreases noise, and the accuracy of the forecast is improved for short times (cf. Fig. \ref{fig:demonstration}). ("5-day forecast")
\end{enumerate}

Finally, note that for $h \sim 500$ days all estimators based on the expOU model lose any information included in $\hat Y(t)$, and converge to the same error level. This is consistent with real data (cf. Sec. \ref{sec:predpow}).

\section{Application to stock market data}
\label{sec:app}

In this section we present an application of the method to actual stock market data. We analyze the Dow Jones Industrial Average (DJIA) index in the period $01/01/1900$--$05/26/2006$, a total of $29,038$ days. In order to work with 
zero-mean returns, the mean return was subtracted from the actual data. Trading volumes for the index are only available for the period $04/01/1993-05/26/2006$, a total of $3,375$ days.

\subsection{Parameter estimation}
\label{sec:param}

We recall that in the estimation procedure presented here one necessarily needs to assume a theoretical model for the volatility. Having done this, the next step is to estimate the parameters involved in the model chosen. For the expOU model, Eqs. (\ref{eq:expou1})-(\ref{eq:expou2}), these parameters are: $m$, $k$ and $\alpha$. 

Before proceeding further we remark that time increments in real data have a finite size since the market always works on discrete times (for daily data the minimum time increment is $1$ day). Thus, in practice, the (infinitesimal) return variation  $dX(t)=X(t+dt)-X(t)$ corresponds to a (finite) return increment  
$\Delta X(t)=X(t+\Delta t)- X(t)$ where $\Delta t$ is the time step between two consecutive ticks. Also the Wiener differentials $dW(t)$ correspond, in mean square sense \cite{gardiner}, to the increments 
\begin{equation}
\Delta W(t)\approx\varepsilon(t)\sqrt{\Delta t},
\label{delta_wiener}
\end{equation}
where $\varepsilon(t)$ is a Gaussian process with zero mean and unit variance \cite{gardiner}. In the present case our time step has a fixed width and is equal to $\Delta t=1$ day. 

Coming back to the estimation of parameters, we show in Appendix \ref{appB} that
\begin{equation}
\ln m\approx (\gamma+\ln 2)/2+\left\langle\ln\left(|\Delta X|/\sqrt{\Delta t}\right)\right\rangle,
\label{log_m}
\end{equation}
where $\gamma=0.5772\cdots$ is the Euler constant. Taking into account that the third summand can be evaluated from data we see that Eq. (\ref{log_m}) provides a direct estimation of $m$. 

\begin{figure}[ptb]
\centerline{\includegraphics[height=180pt]{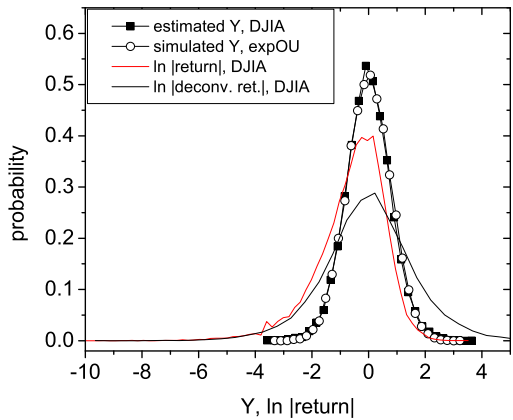}}
\caption{A comparison of estimates of log-volatility for Dow Jones. Black boxes (``estimated $Y$, DJIA'') represent our maximum likelihood method applied to empirical data. Empty circles (``simulated $Y$, expOU'') represent the distribution of the simulated sample path of the log-volatility, assuming that $Y(t)$ follows the expOU model. We also plot the empirical distributions of two estimates: the red line (``ln [return], DJIA'') was obtained through Eq. (\ref{eq:dxabs}) and the black line (``ln [deconv. ret], DJIA'') via Eq. \eqref{eq:deconv}.}
\label{fig:distribution}
\end{figure}

On the other hand, if the expOU model is appropriate then the empirical estimate $\hat{Y}(t)$ of the hidden volatility $Y(t)$ should also be a Gaussian process with a stationary distribution of zero mean and variance given by $\beta=k^2/2\alpha$ [see Eq. (\ref{variance})]. As shown in Fig. \ref{fig:distribution}, if one takes $\beta=0.61\pm 0.05$, the distribution of $\hat{Y}(t)$ is Gaussian and coincides with the theoretical distribution of $Y(t)$ given by Eq. (\ref{eq:p}). The assumption of a Gaussian distribution for our estimate is robust and it holds for a wide range of parameters.

To fully specify the model we have to obtain the parameter $\alpha$. We have chosen the value found in Ref. \cite{masoliver.qf} which was obtained in order to capture the long range correlations of the volatility, at least up to $500$ days. The model is able to provide the appropriate long range behavior with an infinite sum of exponentials but it is also true that it does not provide a pure power law decay like the models in Refs.~\cite{bacry,bns}. In any case, we have seen in Ref. \cite{masoliver.qf} that at least for daily data our approach is satisfactory. Our final set of parameter estimates is thus 
$$
m = (7.5 \pm 0.5) \times 10^{-3}\ \mathrm{days}^{-1/2}, 
$$
$$
\alpha = (1.82 \pm 0.03)\times 10^{-3}\ \mathrm{days}^{-1},
$$
and 
$$
k = (4.7 \pm 0.3) \times 10^{-2}\ \mathrm{days}^{-1/2}.
$$
The errors were determined based on Fig. \ref{fig:distribution}, similarly to the error of $\beta$. In this parameter range the distributions of the estimated $\hat Y(t)$ and simulated $Y(t)$ series agree well. The results reported throughout this paper are insensitive to the misspecification of these parameters, even beyond these error bars (see also Appendix~\ref{app-proc}).

The distributions of log-volatility are compared in Fig. \ref{fig:distribution} for four cases: our maximum likelihood procedure applied to Dow Jones, a simulation of expOU, the simple estimate of Eq. \eqref{eq:deconv} and the deconvolution procedure introduced in Ref. \cite{masoliver.qf} which can be written as
\begin{equation}
\hat Y_\textrm{decon}(t) = \ln \frac{1}{m} \left|\frac{dX(t)}{\widehat{d W_1}(t)}\right|,
\label{eq:deconv}
\end{equation}
where $W_1(t)$ is a simulation of the Wiener process. Note that this deconvoluted log-return estimator has indeed a Gaussian distribution but with a larger variance as hinted in Fig. \ref{fig:distribution} in view of its wider density. We finally mention that the estimate for the log-volatility $\ln(\sigma/m)\approx\ln|dX|$ given by Eq. (\ref{eq:dxabs}) shows a non Gaussian and biased distribution as was also reported in Ref. \cite{masoliver.qf}. This suggests that 
$\hat Y_\mathrm{decon}$ is an appropriate ``null model'' to contrast with $\hat Y$. Both quantities are generated by dividing $dX(t)$ by the increments of a realization of the Wiener process and then taking the logarithm of the absolute value of this ratio. The difference lays in the fact that $\hat Y$ takes the realizations that satisfy a maximum likelihood requirement while $\hat Y_\mathrm{decon}$ takes a Wiener process realization that is purely random. In such a way, our method keeps the divisor correlated with $dX$ and, as we will see next, it seems to conserve clustering and memory effects in the log-volatility time series.

\begin{figure*}[ptb]
\centerline{\includegraphics[height=180pt]{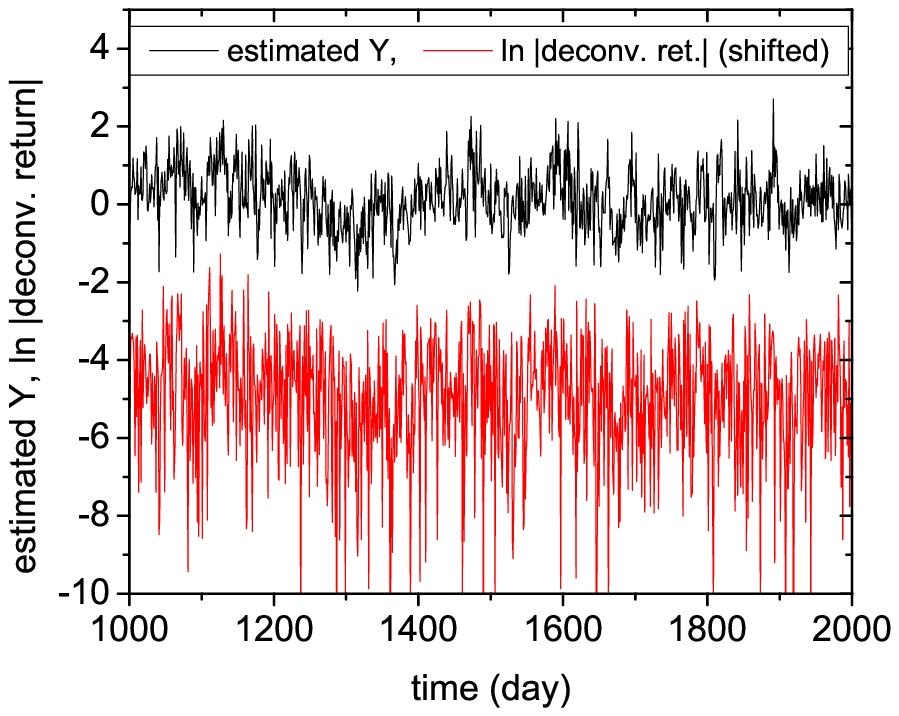}
\includegraphics[height=180pt]{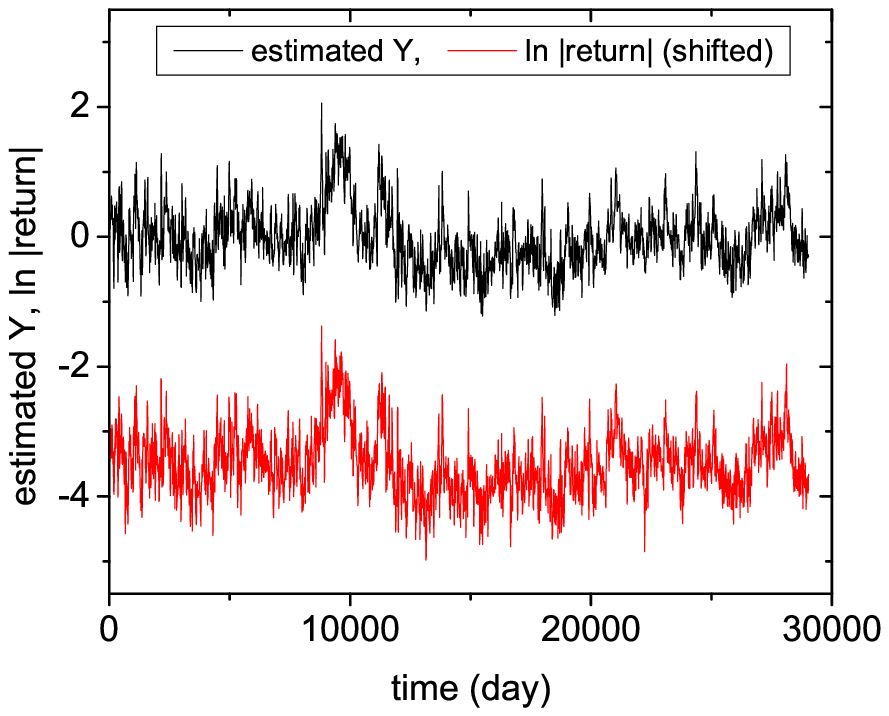}}
\caption{[Left] A comparison between the estimate $\hat Y(t)$ and $\hat Y_\textrm{decon}(t)$ for a typical $1000$-day period of Dow Jones. These curves were not smoothed in order to show the substantial reduction of both the noise level and the asymmetry in $\hat Y(t)$ compared to 
$\hat Y_\textrm{decon}(t)$ with a random approximation of the noise term $dW$. [Right] The estimate $\hat Y(t)$ and the logarithm of absolute return variations for the whole sample of Dow Jones.}
\label{fig:demonstrationDJIA}
\end{figure*}

\subsection{Clustering and the estimated volatility}

To support our claim that the technique presented is powerful enough in filtering the noise $dW_1$ out of returns we will give a visual comparison based on the following qualitative experiments. 

Figure \ref{fig:demonstrationDJIA} [Left] displays a comparison over a $1000$-day time interval. One can observe there that the noise level in $\hat Y$ is substantially smaller than in $\hat Y_\mathrm{decon}$, as also inferred from Fig. \ref{fig:distribution}. 

In order to show that such a correlation is responsible for suppressing large fluctuations in the ratio, we can perform a second experiment. Thus in Fig. \ref{fig:demonstrationDJIA} [Right] we see a comparison between the logarithm of absolute returns variations, $\ln |dX|$, and the estimated volatility $\hat Y$. Note that the proper clustering of volatility becomes clearly visible.

\subsection{A comparison with trading volume}

\begin{figure}[ptb]
\centerline{\includegraphics[height=180pt]{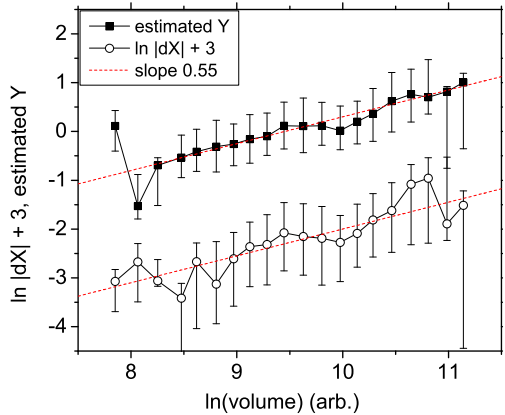}}
\caption{Logarithm of daily absolute return and estimated log-volatility $\hat Y$ as the function of daily volume. Days with similar volumes were binned for better visibility. The symbols represent the medians, and the error bars the $25-75\%$ quantiles in the bins.}
\label{fig:volumedep}
\end{figure}

The hidden nature of the volatility process has been addressed by several authors \cite{yakovenko,andersen,brock}. For instance, Ref. \cite{yakovenko} suggests that, instead of the volatility, a good estimate would be the square root of the daily trading volume. That is, 
\begin{equation}
{\rm M}[\sigma(t)]\propto V(t)^\alpha,
\label{eq:volume}
\end{equation}
where $\alpha = 0.5$ and ${\rm M}[\cdot]$ denotes the median. In Fig. \ref{fig:volumedep} we show evidence that supports this assumption. Again, the first estimation for the volatility is $|dX(t)|$. Therefore, we regress $\ln|dX(t)|$ versus $\ln V(t)$ as shown in Fig. \ref{fig:volumedep}. In the same figure we also present the regression between the maximum likelihood estimate $\hat{Y}(t)$ and $\ln V(t)$ which appears to be less noisy than the former regression in accordance with the smaller variance of $\hat{Y}(t)$ compared to that of 
$\ln|dX(t)|$. Nevertheless, the exponent $\alpha\approx 0.55$ is the same for both regressions.  There have been similar \cite{gabaix.powerlaw}, albeit controversial \cite{farmer.whatreally,eisler.sizematters}, findings for the price impact of single transactions. However, Eq. \eqref{eq:volume} does not yet imply that volatility is proportional to volume, only that \emph{its typical value is} ({\it i.e.}, the median). Fluctuations around the average behavior due to changes in liquidity might have a key role in the process \cite{farmer.whatreally}.

\subsection{The predictive power of volatility\label{sec:predpow}}

\begin{figure}[ptb]
\centerline{\includegraphics[height=240pt]{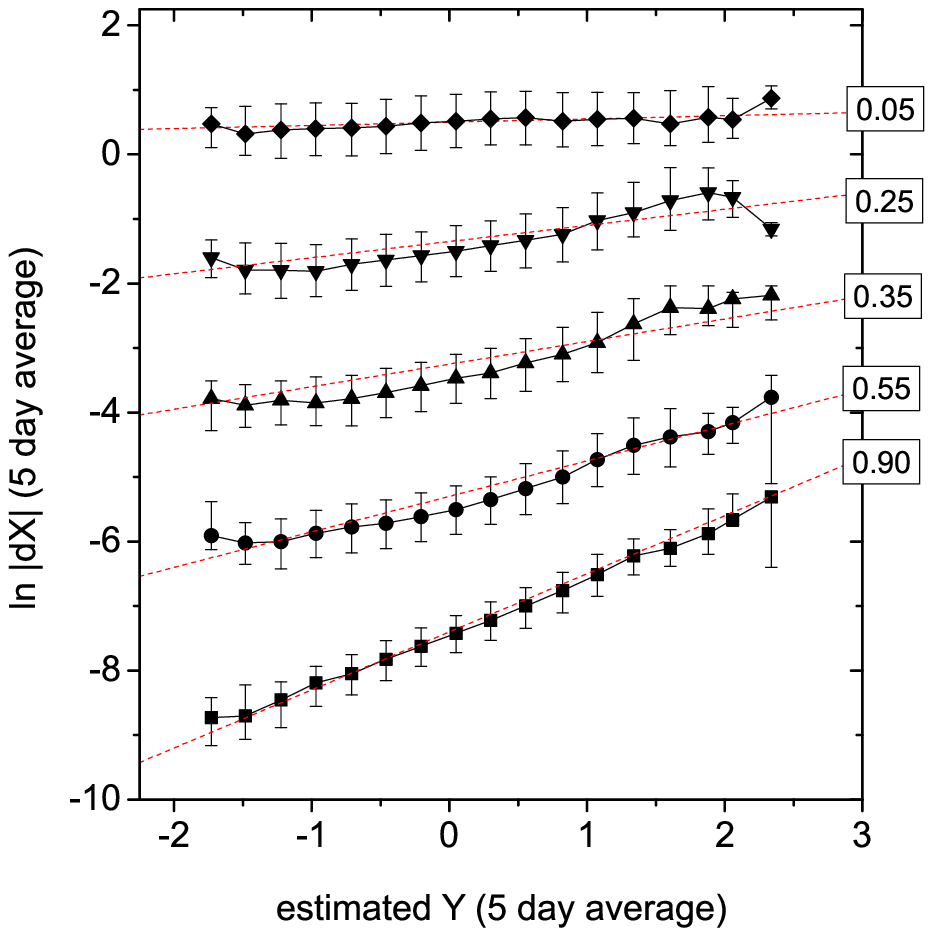}}
\caption{The proportionality between the estimated volatility and of the logarithm of absolute return variations. In order to decrease noise, 5-day moving averages have been used. Numbers on the right indicate the slopes of the corresponding regression lines. Time shifts from bottom to top: $h=0$ days ($\blacksquare$), $5$ days ($\CIRCLE$), $20$ days ($\blacktriangle$), $100$ days ($\blacktriangledown$), $1000$ days ($\blacklozenge$). Days with similar absolute returns were binned for better visibility. The symbols represent the medians, and the error bars the $25-75\%$ quantiles in the bins.}
\label{fig:blockhorizon}
\end{figure}

From Eqs. (\ref{eq:expou1})-(\ref{eq:expou2}), we know that for the expOU model simple relationship can be given between $\ln|dX(t)|$ and $Y(t)$:
$$
\ln|dX(t)|=\ln(m|dW_1(t)|)+Y(t).
$$
Therefore, the conditional median of $\ln|dX(t)|$ given $Y(t)$ is
\begin{equation}
{\rm M}\biggl[\ln |dX(t)|\biggl|\biggr.Y (t) \biggr] = \mathrm{const.}+ Y(t).
\label{median}
\end{equation}
We point out that this relationship implies some degree of predictability of the absolute changes in return through their median, if one knows the current value of the log-volatility $Y(t)$. We test Eq. (\ref{median}) for real data and with $Y(t)$ replaced by its estimate $\hat{Y}(t)$. As shown by the bottom curve of 
Fig. \ref{fig:blockhorizon}, the slope of the linear regression between ${\rm M}\bigl[\ln |dX(t)|\bigl|\bigr.Y\bigr]$ and $\hat Y(t)$ is not equal to $1$ -- as would have been implied by Eq. (\ref{median}) -- but $0.9$ which still suggests strong predictive power. 

Recall that the minimum time step of the empirical data used is $1$ day. Hence, Eq. (\ref{median}) implies the prediction of tomorrow's return based on today's volatility and return. We now want to extend the prediction horizon. To this end we generalize Eq. (\ref{median}) and propose the following ansatz:
\begin{equation}
{\rm M}\biggl[\ln|dX(t+h)|\biggl|\biggr.\hat Y(t) \biggr] = \mathrm{const.}+\gamma(h)\hat Y(t),
\label{median2}
\end{equation}
where $h=0,1,2,\cdots$. In Fig. \ref{fig:blockhorizon} we test this ansatz for several values of the horizon: $h=0, 5, 20, 100$ and $1000$ days. We find that the slope $\gamma(h)$ is a decreasing function from the value $\gamma(0)=0.9$ to practically zero when $h=1000$ days which means a complete loss of memory. Note that, when $h=100$ trading days we have $\gamma=0.25$ still implying a slight degree of prediction after approximately five months, which is of the same order of magnitude than the DJIA characteristic time scale, $1/\alpha\sim 500$ days, for the relaxation of the volatility \cite{masoliver.qf}.  

\section{Conclusions}
\label{sec:conc}

The volatility is a crucial quantity for financial markets since it provides a measure of the amplitude of price fluctuations. Traders try to follow carefully the level of volatility because it gives the perception of the risk associated with any asset. Although volatility was originally conceived to be the diffusion coefficient of the price return random walk, there is compelling evidence not to consider it a constant, but a subordinated random process. The framework is analogous to that of random diffusion processes which have been applied to a large variety of phenomena in statistical mechanics and condensed matter physics.

The main obstacle to get a better knowledge of the volatility's nature is that it is not directly observed. In fact, this is precisely the motivation behind the present research. Our main objective has been to develop a tool which visualizes the sample path of volatility. The procedure derives a maximum likelihood estimate assuming that the volatility is a hidden Markov process. To do so, one needs also to assume a specific model for the volatility. We have chosen a class of two dimensional diffusions commonly known as stochastic volatility models, where the volatility acts as a diffusion particle trapped in a potential well. We have focused on the expOU model and obtained promising results, especially for three reasons: (i) the model is computationally feasible; (ii) its parameters can be easily obtained and fit the data reasonably well; and (iii) the distribution of the estimated volatility is log-normal, which is consistent with the assumed expOU model.

We have shown for the Dow Jones index daily data that the sample path of our estimated volatility improves other estimates. We have compared our estimation with a rather typical one which identifies volatility with absolute return changes. Our estimation is able to remove the existing bias in the stationary distribution of volatility while still preserving the clustering in volatility time series. We have also studied the estimate of volatility that deconvolutes the return by the simulation of a random Wiener path \cite{masoliver.qf}. This last procedure also provides a Gaussian distribution for the log-volatility, albeit the distribution has too fat tails and pays the price of losing clustering and memory in the volatility time series. Our new procedure is in fact a more sophisticated variant of this estimate since it filters out Wiener realizations via maximum likelihood. The estimate drastically reduces the noise in the volatility path thus preserving data clustering. In this way we have thus proposed an alternative methods to those already provided by mathematical finance literature (see e.g. \cite{bns,andersen2,genon,elerian,bns2,barucci,genon2,genon3,griffin,jong,morimune,omori,reno}). It should however be left for a future research an accurate comparison between our method and others. We may even look for a way to implement parameter estimation in our method.

The median of the estimated volatility has also been related to trading volume by the power-law expression $M[\sigma]\propto V^{0.55}$. A link between volatility and trading volume has been previously mentioned in different studies however our estimate is again capable to provide a less noisy regression. We must, indeed, stress the fact that this does not imply that volatility is proportional to a power of the volume, but only that its typical value is and that fluctuations around the average might play an important role.

We have also seen that current returns are proportional to the estimated volatility, as otherwise expected. However, the main novelty is that we have observed how future returns are proportional to current volatility and their predictability diminishes monotonically with the number of time steps ahead. This last finding implies that our estimation method can be applied to predict the size of future returns with the knowledge of current volatility.

As a final remark we stress the fact that the technique herein presented can be applied to a variety of physical phenomena besides finance. One typical problem of this sort is provided by the Brownian motion inside a field of force in which inertial effects are not negligible \cite{masoliver93}. In this situation the dynamics of the particle is described by a two dimensional diffusion process $(X(t),V(t))$ representing the position and the velocity of the Brownian particle. The maximum likelihood technique might provide a reliable estimate of the velocity in the case that, for instance, the only accessible experimental measures are the positions of the particle at wide time steps, so that a measure of the velocity -- which implies the knowledge of two very close positions -- is too noisy and unreliable.  

\section*{Acknowledgments}

ZE is grateful to J\'anos Kert\'esz for his support and to the Universitat de Barcelona for its hospitality during his visit at the Departament de Fisica Fonamental; also support by OTKA T049238 is acknowledged. JP and JM acknowledge support from Direcci\'on General de Investigaci\'on under contract No. FIS2006-05204.

\appendix

\section{Derivation of the likelihood function}
\label{app:proof}

In order to make notations more compact, in this Appendix the time dependence of the stochastic processes is mostly indicated as a lower index. A generic stochastic volatility model is defined as
\begin{eqnarray}
&&dX_t = f(Y_t)dW_{1}(t), \label{a1} \\
&&dY_t = -g(Y_t)dt+h(Y_t)dW_{2}(t).
\label{a2}
\end{eqnarray}

To explain the procedure it is more convenient to work with the discrete time version of the model. To this end, suppose that $\Delta t$ is a small time step and that the driving noises in Eqs. (\ref{a1})-(\ref{a2}) can be approximated by ({\it cf} Eq. (\ref{delta_wiener}))
\begin{equation}
dW_{i}(t) \approx\varepsilon_{i}(t)\sqrt{\Delta t}, \qquad (i=1,2),
\label{a3}
\end{equation}
where $\varepsilon_i(t)$ are independent standard Gaussian processes with zero mean and unit variance. We remark that the approximation (\ref{a3}) has to be understood in mean square sense \cite{gardiner}. The discrete time equations of the model thus read
\begin{eqnarray}
&&X_t-X_{t-\Delta t} = f(Y_{t-\Delta t})\varepsilon_{1}(t-\Delta t)\sqrt{\Delta t}\, \label{a4} \\
&&Y_t-Y_{t-\Delta t} = -g(Y_{t-\Delta t})\Delta t+h(Y_{t-\Delta t})\varepsilon_{2}(t-\Delta t)\sqrt{\Delta t}\;
\label{a5}
\end{eqnarray}
from which we get
\begin{eqnarray}
\label{a6}
&& \varepsilon_{1}(t-\Delta t) = \frac{X_t-X_{t-\Delta t}}{f(Y_{t-\Delta t})\sqrt{\Delta t} },\\
\label{a7}
&& \varepsilon_{2}(t-\Delta t)=\frac{Y_t-Y_{t-\Delta t}+g(Y_{t-\Delta t})\Delta t}{h(Y_{t-\Delta t})\sqrt{\Delta t}}.
\end{eqnarray}
For a given number of realizations, the probability of the set $\{X_\tau,Y_\tau\}$ $(\tau=t-\Delta t,t- 2\Delta t,\dots,t-s)$ can be easily obtained, as we will see next.

Let us denote the set of realizations as $\{\mathbf X, \mathbf Y\}$. Then the Markov property of the process ensures that one can decompose the joint probability density function (pdf) of this set as a chain of products between conditional probability densities. In consequence, the pdf of the whole sample path can be written as
\begin{widetext}
\begin{equation}
\mathbb P(\{\mathbf X, \mathbf Y\})=
\mathbb P(X_{t-s},Y_{t-s})\prod_{\tau=t-s}^{t-\Delta t}\mathbb P(X_{\tau},Y_{\tau}\vert X_{\tau-\Delta t},Y_{\tau-\Delta t}),
\label{a8}
\end{equation}
\end{widetext}
where the first term, $\mathbb P(X_{t-s},Y_{t-s})$, corresponds to the initial realizations of $X$ and $Y$ $s/\Delta t$ time steps far from the present time $t$; all the remaining terms of the form $\mathbb P(X_{\tau},Y_{\tau}\vert X_{\tau-\Delta t},Y_{\tau-\Delta t})$ are the conditional pdf's for transitions between consecutive steps: 
$$
(X_{\tau-\Delta t},Y_{\tau-\Delta t})\longrightarrow (X_{\tau},Y_{\tau}).
$$
The logarithm of Eq. (\ref{a8}) is
\begin{widetext}
\begin{equation}
\ln \mathbb P(\{\mathbf X, \mathbf Y\})= \ln \mathbb P(X_{t-s},Y_{t-s}) +
\sum_{\tau=t-s}^{t-\Delta t}\ln \mathbb P(X_{\tau},Y_{\tau}\vert X_{\tau-\Delta t},Y_{\tau-\Delta t}).
\label{a9}
\end{equation}
\end{widetext}

On the other hand from Eqs. (\ref{a4})-(\ref{a5}) we realize that
$$
\mathbb P(X_\tau,Y_\tau\vert X_{\tau-\Delta t},Y_{\tau-\Delta t})=|J|\mathbb P(\varepsilon_1(\tau-\Delta t),\varepsilon_2(\tau-\Delta t)),
$$
where $|J|$ is the Jacobian of the transformation $(X_\tau,Y_\tau)\longrightarrow(\varepsilon_1(\tau-\Delta t),\varepsilon_2(\tau-\Delta t))$ defined by Eqs. (\ref{a4})-(\ref{a5}), that is, 
$$
|J|=\frac{1}{f(Y_{\tau-\Delta t})h(Y_{\tau-\Delta t})\Delta t}.
$$
But $\varepsilon_1$ and $\varepsilon_2$ are independent standard Gaussians, hence
$$
\mathbb P(\varepsilon_1,\varepsilon_2)=(1/2\pi)
\exp\left[(\varepsilon_{1}^2+\varepsilon_{2}^2)/2\right],
$$
whence
\begin{widetext}
\begin{equation}
\mathbb P(X_\tau,Y_\tau\vert X_{\tau-\Delta t},Y_{\tau-\Delta t}) = \frac{1/(2\pi\Delta t)}
{f(Y_{\tau-\Delta t})h(Y_{\tau-\Delta t})}
\exp\left[-\frac{\varepsilon_{1}^2(\tau-\Delta t)+\varepsilon_{2}^2(\tau-\Delta t)}{2}\right].
\label{a10}
\end{equation}
\end{widetext}
Substituting Eqs. (\ref{a6})-(\ref{a7}) into this equation and the result into Eq. (\ref{a9}) we finally get
\begin{eqnarray}
\ln \mathbb P(\{\mathbf X, \mathbf Y\})&=&-\frac{s\ln(2\pi\Delta t)}{\Delta t}-\sum_{\tau=t-s}^{t-\Delta t}[\ln f(Y_{\tau-\Delta t})+ \ln h(Y_{\tau-\Delta t})] \nonumber \\
&+&\ln \mathbb P(X_{t-s}, Y_{t-s})
-\frac{1}{2}\sum_{\tau=t-s}^{t-\Delta t}\left[\frac{X_\tau-X_{\tau-\Delta t}}{f(Y_{\tau-\Delta t})\Delta t}\right]^2 \Delta t \nonumber \\
&-&\frac{1}{2}\sum_{\tau=t-s}^{t-\Delta t}\left[\frac{Y_\tau-Y_{\tau-\Delta t}}{h(Y_{\tau-\Delta t})\Delta t}+
\frac{g(Y_{\tau-\Delta t})}{h(Y_{\tau-\Delta t})}\right]^2\Delta t.
\label{a11}
\end{eqnarray}

Let us briefly explain the origin of some of these contributions. The first summand comes from the normalization constant of the Gaussian distribution~(\ref{a10}). It appears in every conditional probability density and this is the reason for the factor $s/\Delta t$, which is the number of time steps between $t-s$ and $t$. The resulting term does not depend on the realization, so that we can neglect it for a maximization with respect to $\mathbf Y$. The term also goes to $-\infty$ in the $\Delta t \rightarrow 0$ limit, which means that any individual realization has a probability measure zero. 

The second summand is mostly the sum of the Jacobian transformations of each transition probability and depends on $Y$. Stochastic volatility models typically assume that these $f$ and $g$ [cf. Eqs. (\ref{a1}) and (\ref{a2})] are continuous and monotonically increasing functions or even constants. For instance, in the expOU model we have $f(x)=m\exp(x)$ and $g(x)=k$. Because of this, we will also neglect this term in the maximization procedure. The contribution shifts the maximum at the excessive cost of adding more noise to the numerical computations. We can however look at the situation from another point of view. Ignoring this term is equivalent to omitting the Jacobian transformation between the two probability density measures [cf. Eq.(\ref{a10})]. In this way, we are stating that what we are really going to maximize is the probability of the realization of $\varepsilon_1(t)$ and $\varepsilon_2(t)$ --instead of $Y(t)$-- in terms of the past history of the process expressed by $\{\mathbf X, \mathbf Y\}$. 

The term $\ln\mathbb P(X_{t-s},Y_{t-s})$ is fixed by the initial conditions of the process. If we assume a known initial return $X$ -- which can be set to zero -- and take a random $Y_{t-s}$ following the stationary distribution $\mathbb P_\mathrm{st}(Y_{t-s})$ given by Eq.~(\ref{eq:p}), then $\mathbb P(X_{t-s},Y_{t-s})=\delta(X_{t-s}-X)\ P_\mathrm{st}(Y_{t-s})$ and hence
\begin{equation}
\ln \mathbb P(X_{t-s},Y_{t-s})=\ln \mathbb P_\mathrm{st}(Y_{t-s})+\ln\delta(X_{t-s}-X).
\end{equation}
Had we taken another initial condition, the technique would have given equivalent results (we have checked this by using several initial distributions). For this reason and in order to improve the convergence of the maximum likelihood estimate we have neglected also this contribution.

We therefore write
\begin{eqnarray}
\ln \mathbb P(\{\mathbf X, \mathbf Y\})&\simeq&
-\frac{1}{2}\sum_{\tau=t-s}^{t-\Delta t}\left[\frac{X_\tau-X_{\tau-\Delta t}}{f(Y_{\tau-\Delta t})\Delta t}\right]^2 \Delta t \nonumber\\ 
&-&\frac{1}{2}\sum_{\tau=t-s}^{t-\Delta t}\left[\frac{Y_\tau-Y_{\tau-\Delta t}}{h(Y_{\tau-\Delta t})\Delta t}+
\frac{g(Y_{\tau-\Delta t})}{h(Y_{\tau-\Delta t})}\right]^2\Delta t+\cdots
\label{a12}
\end{eqnarray}
We can represent this equation in the continuous time framework if $\Delta t$ is sufficiently small and if $f(x)$, $h(x)$ and $g(x)$ are continuous. In such a case, Eq.~(\ref{a12}) yields the result given in Eq.~(\ref{eq:logp}). 

\section{Robustness of the procedure}
\label{app-proc}

\begin{figure}[tb]
\centerline{\includegraphics[height=240pt]{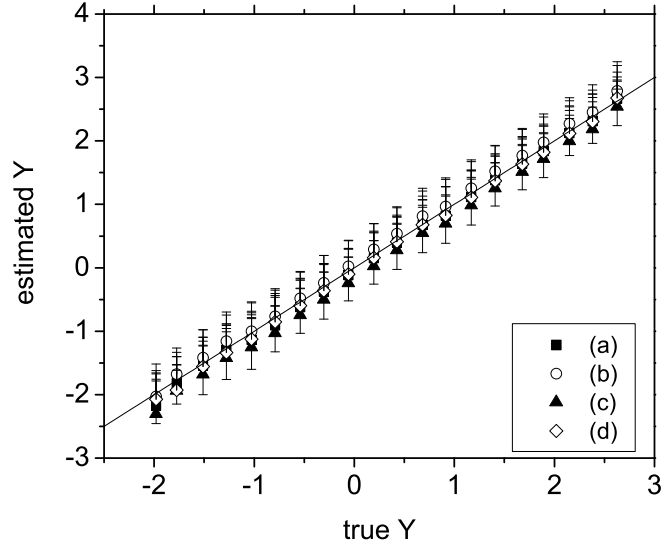}}
\caption{Estimated log-volatility $\hat Y$ as a function of the actual log-volatility $Y$ taken from $29038$ simulations of the expOU model in four ways. Reconstruction uses the last $s=10$ values of returns, and $I=10^5$ iterations. The error bars represent the $25\%$ and $75\%$ quantiles of the distribution estimated volatility.}
\label{fig:app}
\end{figure}

In order to show that our volatility estimation procedure is robust, we carry out four measurements along the lines of Sec.~\ref{sec-perf}. We generate artificial time series of the same length as the DJIA dataset, with the same parameters as therein: $m = (7.5 \pm 0.5) \times 10^{-3}\ \mathrm{days}^{-1/2}$, $\alpha = (1.82 \pm 0.03)\times 10^{-3}\ \mathrm{days}^{-1}$, and $k = (4.7 \pm 0.3) \times 10^{-2}\ \mathrm{days}^{-1/2}$.

Then we reproduce Fig.~\ref{fig:reconstruction} in several variations:

\begin{enumerate}[(a)]
\item In a way identical to the paper as shown in Fig.~\ref{fig:reconstruction}.
\item Data generated as in (a), but for reconstructing $\hat Y$ we use the above parameters \emph{minus} twice the error specified above.
\item Data generated as in (a), but for reconstructing $\hat Y$ we use the above parameters \emph{plus} twice the error specified above.
\item In a way identical to (a), but the data is detrended by the Monte Carlo sample mean $\mu$. We however remind that the process is driftless as given by Eq.~(\ref{1r}). In this way we want to check whether detrending causes a systematic bias in the estimation.
\end{enumerate}

The results are given in Fig.~\ref{fig:app}. The plot shows, that either option gives the same width of the $\hat Y$ distribution, and the bias introduced by either specification error is small. We have also run similar simulations but with very different parameters values and the procedure still provides consistent results similar to those given by Fig.~\ref{fig:app}. 

\begin{figure}[tb]
\centerline{\includegraphics[height=240pt]{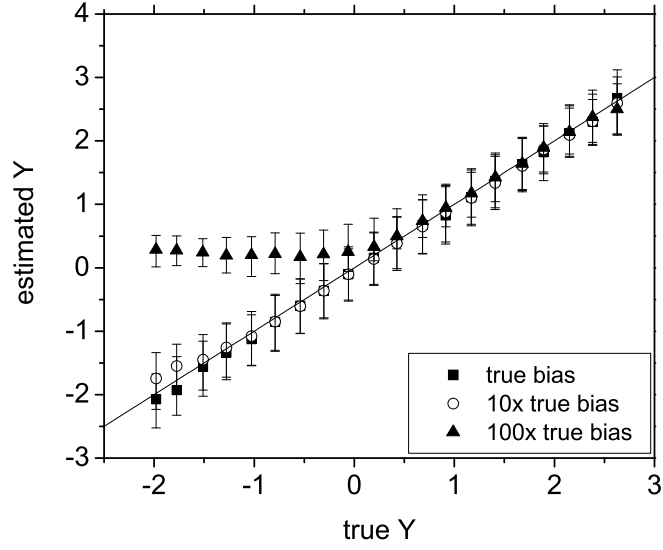}}
\caption{Estimated log-volatility $\hat Y$ as a function of the actual log-volatility $Y$ taken from $29038$ simulations of the expOU model in three different levels of "detrending error". Reconstruction used the last $s=10$ values of returns, and $I=10^5$ iterations. The error bars represent the $25\%$ and $75\%$ quantiles of the distribution estimated volatility. Note that there only exists a clear deviation when the error is 100 times the true bias.}
\label{fig:app1}
\end{figure}

Finally, a more detailed test along the lines of point (d) above can be performed (that is: a test on wether the detrending causes a bias in our estimation procedure). We thus have repeated the test (d) also including errors $10\mu$ and $100\mu$. The procedure tolerates up to $10$ times more detrending error than expected in the real dataset. Only, when we have about $100$ times the error, the reconstruction shows a strong upward bias for low volatility periods as shown in Fig. \ref{fig:app1}. All these results imply that possible errors in the detrending procedure could affect our procedure only when any wrong specification of the drift is far outside the error domain involved in our DJIA data set.

\section{Derivation of Eq. (20)}
\label{appB}

We start from Eq. (\ref{eq:expou1}) which we write in the approximate form
$$
\Delta X(t)\simeq me^{Y(t)}\Delta W(t),
$$
thus $\ln|\Delta X(t)|=m+Y(t)+\ln|\Delta W_1(t)|$ and, taking into account that $\langle Y(t)\rangle=0$, we have
\begin{equation}
\langle\ln|\Delta X(t)|\rangle\simeq\ln m+\langle\ln|\Delta W_1(t)|\rangle.
\label{b1}
\end{equation}

On the other hand, we know that $\Delta W_1(t)\approx \varepsilon\sqrt{\Delta t}$, where $\varepsilon$ is a standard Gaussian variable ({\it cf} Eq. (\ref{delta_wiener})). Hence
$$
\langle\ln|\Delta W_1(t)|\rangle\approx\langle\ln|\varepsilon|\rangle+(\ln \Delta t)/2.
$$
But
$$
\langle\ln|\varepsilon|\rangle=\frac{1}{\sqrt{2\pi}}\int_{-\infty}^{\infty}e^{-\varepsilon^2/2}\ln|\varepsilon|d\varepsilon,
$$
which, after a simple change of variables inside the integral, yields \cite{gradstein,mos}
\begin{widetext}
$$
\langle\ln|\varepsilon|\rangle=\frac{1}{2\sqrt{2\pi}}\int_{0}^{\infty}x^{-1/2}e^{-x/2}\ln x dx=
-\sqrt{\pi/2}(\gamma+\ln 2),
$$
\end{widetext}
where $\gamma=0.5772\cdots$ is the Euler constant. Therefore,
\begin{equation}
\langle\ln|\Delta W_1(t)|\rangle\approx(\ln \Delta t)/2-(\gamma+\ln 2)/2.
\label{b3}
\end{equation}
Substituting Eq. (\ref{b3}) into Eq. (\ref{b1}) proves Eq. (\ref{log_m}).

\bibliographystyle{apsrev}
\bibliography{Volatility-resub1}

\end{document}